\documentclass{opticajnl}
\journal{opticajournal} 
\setboolean{shortarticle}{true}
\usepackage{subcaption}
\usepackage{gensymb}
\usepackage{hyperref}
\usepackage{ulem}

\title{Indium tin oxide combined with anti-reflective coatings with high transmittance for wavelengths $<$ 400 nm}

\author[1,*]{Erik Jansson} 
\author[2,$\dag$]{Volker Scheuer} 
\author[1]{Elena Jordan} 
\author[2]{Konstantina Kostourou}
\author [1,3,4]{Tanja  E. Mehlst{\"a}ubler}
\affil[1]{Physikalisch-Technische Bundesanstalt, Bundesallee 100, 38116 Braunschweig, Germany}
\affil[2]{NANEO Precision IBS Coatings GmbH Heuriedweg 31a, 88131 Lindau Germany}
\affil[3]{Institut f{\"u}r Quantenoptik, Leibniz Universit{\"a}t Hannover, Welfengarten 1, 30167 Hannover, Germany}
\affil[4]{Laboratorium f{\"u}r Nano- und Quantenengineering, Leibniz Universit{\"a}t Hannover, Schneiderberg 39, 30167 Hannover, Germany}

\affil[*]{erik.jansson@ptb.de}
\affil[$\dag$]{Deceased April 16, 2024.}

\begin{abstract} 
	The transparent and conductive properties of indium tin oxide (ITO) thin films, make them an attractive coating for optically integrated ion traps. However, the relatively low transmittance for wavelengths $<$ 400 nm, high scattering and high production temperature limits the usability in trapped-ion-based quantum technologies. Here we present ITO coatings and a combined ITO + anti-reflective (AR) coating system optimized for an ion trap applied using ion beam sputtering (IBS). The coatings feature a high transmittance for wavelengths $<$ 400 nm and additional wavelengths up to 1000 nm, low scattering and low production temperature $<$ 150 $\degree$C. The transmission, reflection and absorption spectra are simulated and the resistance, transmittance and scattering at 370 nm are measured for different ITO coating thicknesses and the ITO + AR coating system. For the ITO + AR coating system a resistance of 115 $\pm$ 5 $\Omega/\Box$, transmittance of 80$\%$ and scattering of 0.012 $\pm$ 0.002$\%$ at 370 nm is achieved. 
\end{abstract}
\setboolean{displaycopyright}{false} 
\begin{document}	

\maketitle

\section{Introduction}
Indium Tin Oxide (ITO) is a transparent, conducting material that can be applied as a thin film coating. It is produced by doping indium oxide, a ceramic material that is transparent in the visible spectrum with the metal tin, creating a conductive coating with a high transmittance in the visible spectrum. The combination of electrical and optical properties of the coating makes it an attractive intermediator between electrical and optical interfaces \cite{ma5040661}. As a result, ITO coatings are used in a wide variety of devices for example solar cells and LCDs \cite{ali_structural_2014,pla_optimization_2003, katayama_tft-lcd_1999}. With the rise of Quantum Technologies (QTs), the interest in ITO has increased since QTs often rely on both electrical and optical properties of materials \cite{kutas_terahertz_2020,ji_highly_2017,niffenegger_integrated_2020,karan_mehta_ion_2023,eltony_transparent_2013}. A promising platform in QTs is ion traps, where the coatings have been used for grounding of optics \cite{niffenegger_integrated_2020, karan_mehta_ion_2023} and as a transparent electrode material \cite{eltony_transparent_2013, shcherbinin_charged_2023}. However, ITO typically has low transmittance < 60$\%$ for wavelengths < 400 nm \cite{zhang_near_2022,kim_highly_2017,guillen_influence_2006,shen_room-temperature_2010}, high surface roughness > 1 nm \cite{hao_thickness_2008,kim_low_2000,sterligov_fabrication_2012,lucas_ito_2006} and high production temperatures \cite{senthilkumar_structural_2010,boycheva_structural_2007,strumpfel_low_2000,ren_key_2022,seki_indiumtin-oxide_2001} > 150 $\degree$C which limits the applications of ITO in ion trap experiments. In ion traps, the ion species determines the wavelengths necessary to control the system and wavelengths $<$ 400 nm are common. Therefore, ITO coatings with high transmittance for these wavelengths are needed. High surface roughness can lead to increased scattering that could impair the state readout of the ions \cite{Hultaker_2003}. Ion trap experiments rely on the detection of the fluorescence light of single ions and scattered light increases the detection noise. Furthermore, ultraviolet (UV) light can charge up materials leading to a displacement of the ions \cite{wang_laser-induced_2011}. A further issue when using ITO for ion traps is the high process or annealing temperatures. As an example, magnetron sputtering often requires a substrate temperature of about 200-300 $\degree$C during the coating process or an annealing temperature of 350 $\degree$C after the coating process \cite{strumpfel_low_2000, sterligov_fabrication_2012, guillen_influence_2006}, which can degrade or damage other materials. This creates a demand for ITO coatings with high transmittance and low scattering especially for wavelengths $<$ 400 nm and low production temperatures.\\
Here we present an ITO coating with high transmittance for wavelengths $<$ 400 nm and additionally high transmittance at other technically required wavelengths up to 1000 nm. The coating is applied using IBS, a technique which uses temperatures that can stay below 100 $\degree$C during the production process. IBS is known to create coating layers with lower optical scattering for oxide materials like Ta$_2$O$_5$ and SiO$_2$ compared to for example e-beam deposition \cite{detlev_ristau_ion_2005,Lyngnes:06}. Here we investigate, if using IBS for the production of ITO layers also achieves low optical scattering. The transmittance of the ITO layers is increased by an antireflection (AR) coating at application requested wavelengths. Here we report the design of the ITO+AR coating and the measurement of the transmittance and optical scattering of the coatings at 370 nm. The achieved properties make these ITO layers particularly suitable for the use in ion traps with ytterbium (Yb$^{+}$), calcium (Ca$^{+}$) or strontium (Sr$^{+}$) ions. In this work, we characterize the ITO coatings for the transition wavelengths of Yb$^{+}$ ions. In particular the coatings are used within the BMBF-project IDEAL \cite{noauthor_integrierte_2024}. The IDEAL project develops an ion trap with integrated gradient index (GRIN) lenses for addressing Yb$^{+}$ ions. The integrated GRIN-lenses are used to focus laser light onto the ions in the trap and to collect the light from the fluorescing ions. In this project, ITO coatings are used for the grounding of GRIN-lenses integrated in the ion trap. 
\section{Ion beam sputtering}
NANEO owes the high quality of its coatings to its manufacturing process, i.e. IBS. No other coating technique allows process parameters such as energy and growth rate to be set precisely and independently from one another. The result of IBS is the generation of very dense and compact coatings. The extremely smooth and homogeneous layers with regard to the optical properties of the product enables the production of outstanding optical interference systems \cite{detlev_ristau_ion_2005,timothy_w_jolly_ion-beam_1993}. Several hundreds of layers can be stacked using this method, thereby transforming the substrates into optical components with complex functionality and the highest quality standards. The advantage of IBS compared to other coating processes lies in the extremely low intrinsic absorption together with very low scatter losses of the layers, paired with a very high long-term stability of their optical properties. Precision coatings are only possible if the measurement technique meets the high requirements of the IBS coating. The precise measurement technology in our IBS system is a development of NANEO. Permanent dispersion and layer thickness measurements accompany the whole production process with a measurement resolution as low as 1 $\mathring{\mathrm{A}}$ or less.
\section{ITO coating process development}
The transfer of laser light to and from an ion trap with integrated optics requires an optical coating with a good electrical conductivity to electrically ground the optical surfaces close to the ions in the trap. Therefore, all coating designs presented in this work are developed with an ITO layer on top of the underlying AR system. For the calculation of all the presented optics, the dispersion of the ITO layers had to be taken into account. Moreover, considering that the absorption edge of ITO starts at around 500 nm (Fig. \ref{fig: A_diff_thickness}), including the spectral absorption progression was also necessary. By varying the beam voltage, the process oxygen backpressure, and the primary ion gas, we managed to develop an ITO coating process, which led to an optimized ITO dispersion. Additionally, a water supply for the control of water backpressure in the coating process was added to a commercial IBS machine \cite{konstantina_kostourou_high_2023}. To obtain the required data, several ITO layers with different thicknesses have been coated. The transmission and reflectance of the coated substrates have been measured with a spectrophotometer in the wavelength range from 330 nm to 1000 nm \cite{konstantina_kostourou_high_2023}. The spectral transmission and reflectance measurements and additionally several laser calorimetric measurements of the absorption have been used to determine the dispersion formula of ITO. 
\section{Calculation of the theoretical spectral curves of the ITO layers}
\begin{figure}[ht!]
	\begin{subfigure}[h]{0.5\linewidth}
		\subcaption{}
		\includegraphics[width=\linewidth]{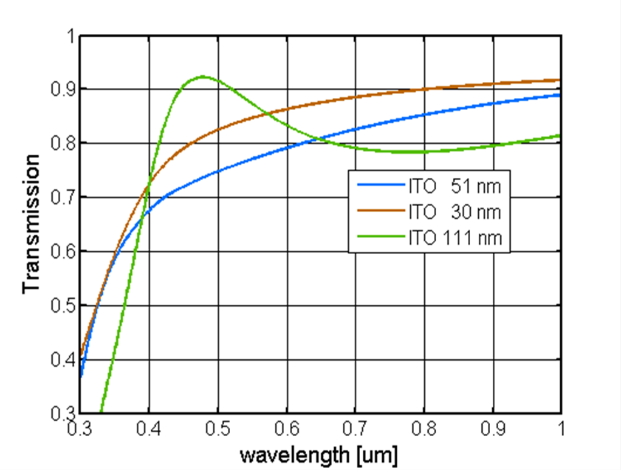}
		\label{fig: T_diff_thickness}	
	\end{subfigure}
	\begin{subfigure}[h]{0.53\linewidth}
		\subcaption{}
		\includegraphics[width=\linewidth]{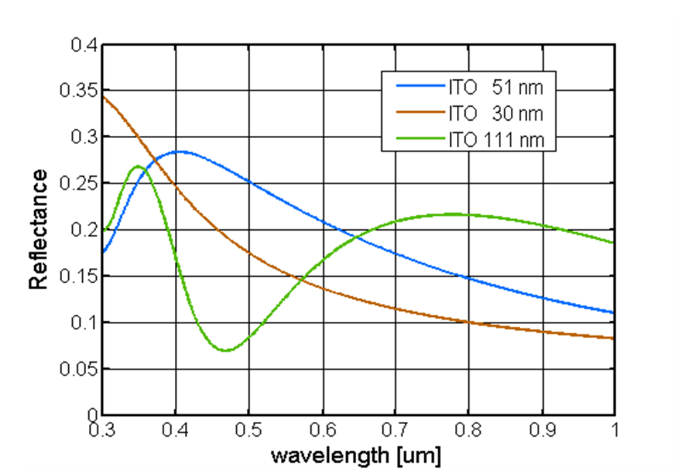}
		\label{fig: R_diff_thickness}	
	\end{subfigure}
\caption{Calculated transmission (\subref{fig: T_diff_thickness}) and  reflectance (\subref{fig: R_diff_thickness}) of single ITO layers of different thicknesses.}
\label{fig: T_R_diff_thickness} 
\end{figure}

\begin{figure}[ht!]
	\begin{subfigure}[h]{0.5\linewidth}
		\subcaption{}
		\includegraphics[width=\linewidth]{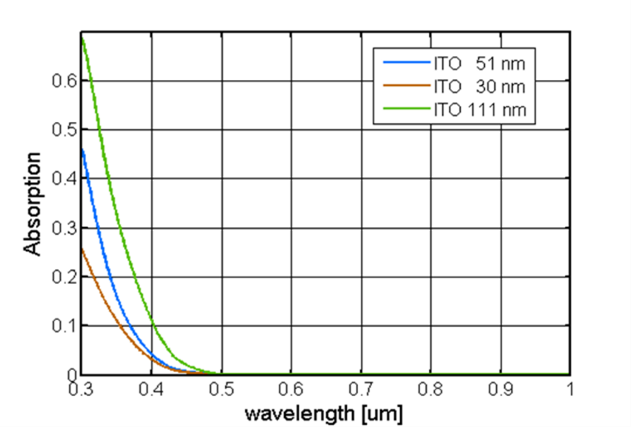}
		\label{fig: A_diff_thickness_a}	
	\end{subfigure}
	\begin{subfigure}[h]{0.53\linewidth}
		\subcaption{}
		\includegraphics[width=\linewidth]{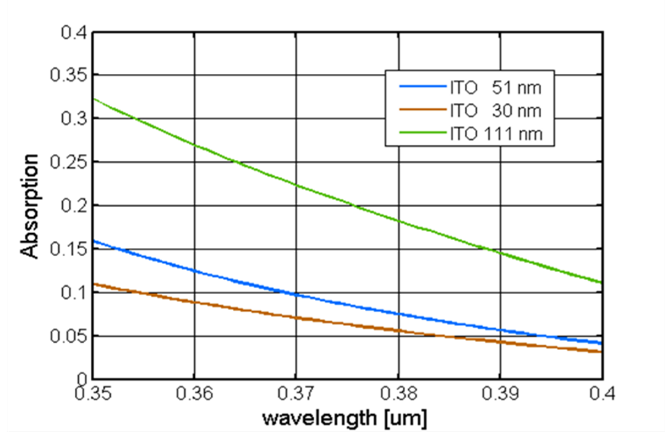}
		\label{fig: A_diff_thickness_b}	
	\end{subfigure}
	\caption{Calculated absorption curves of single ITO layers of different thicknesses at a wavelength range of 300 nm – 1000 nm (\subref{fig: A_diff_thickness_a}) and a zoomed-in region of the same curves at a wavelength range of 350 nm - 400 nm (\subref{fig: A_diff_thickness_b}).}
	\label{fig: A_diff_thickness} 
\end{figure}

\begin{table}[ht!]
	\centering
	\caption{\bf ITO thickness and resistance of the measurement samples set-1.}
\begin{tabular}{ccc}
	\hline
\textbf{Sample}	& \textbf{ITO thickness $\left[nm\right]$} &\textbf{Resistance $\left[ \Omega/\Box \right]$ }  \\
	\hline
ITO30 & 30 & 210 $\pm$ 8 \\
	\hline
ITO51 & 51 & 115 $\pm$ 5\\
	\hline
ITO111 & 111 & 57 $\pm$ 2\\
	\hline
\end{tabular}
\label{tab:thickness_vs_resistance}
\end{table}
Three different ITO layer thicknesses have been coated (later referred to as set-1). The sheet resistance of these layers was measured with a four-point probe system and is given in Table \ref{tab:thickness_vs_resistance}. 
\noindent The theoretical curves for the transmission, reflectance and absorption are calculated using the optimized ITO dispersion. Fig. \ref{fig: T_R_diff_thickness} and \ref{fig: A_diff_thickness} show the results of the calculation for the transmission, reflectance and absorption of the three samples used for the measurements that will be presented in the upcoming paragraphs. These curves have been calculated, taking into consideration both substrate interfaces. For more information, we refer to Fig. \ref{fig:Ex_optical_interfaces} and the corresponding paragraph. The transmission and reflectance at different wavelengths are determined by optical interference and absorption of the ITO layers. 
\section{Optimizing the transmission at specific wavelengths for the ion trap application}
\begin{figure}[ht!]
	\begin{subfigure}[h]{0.5\linewidth}
		\subcaption{}
		\includegraphics[width=\linewidth]{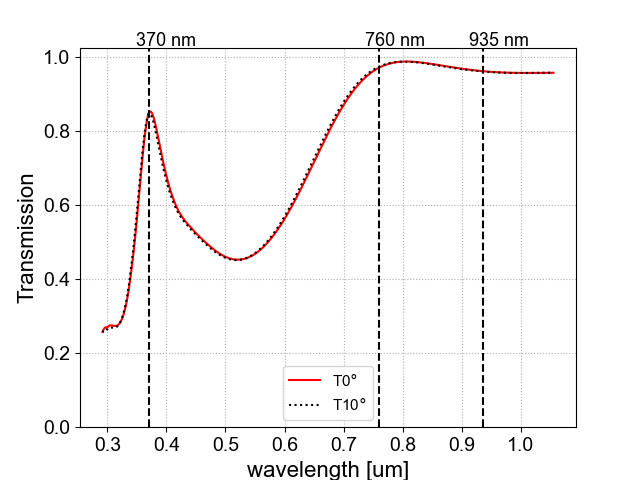}
		\label{fig: T_Z}	
	\end{subfigure}
	\\
	\begin{subfigure}[h]{0.5\linewidth}
		\subcaption{}
		\includegraphics[width=\linewidth]{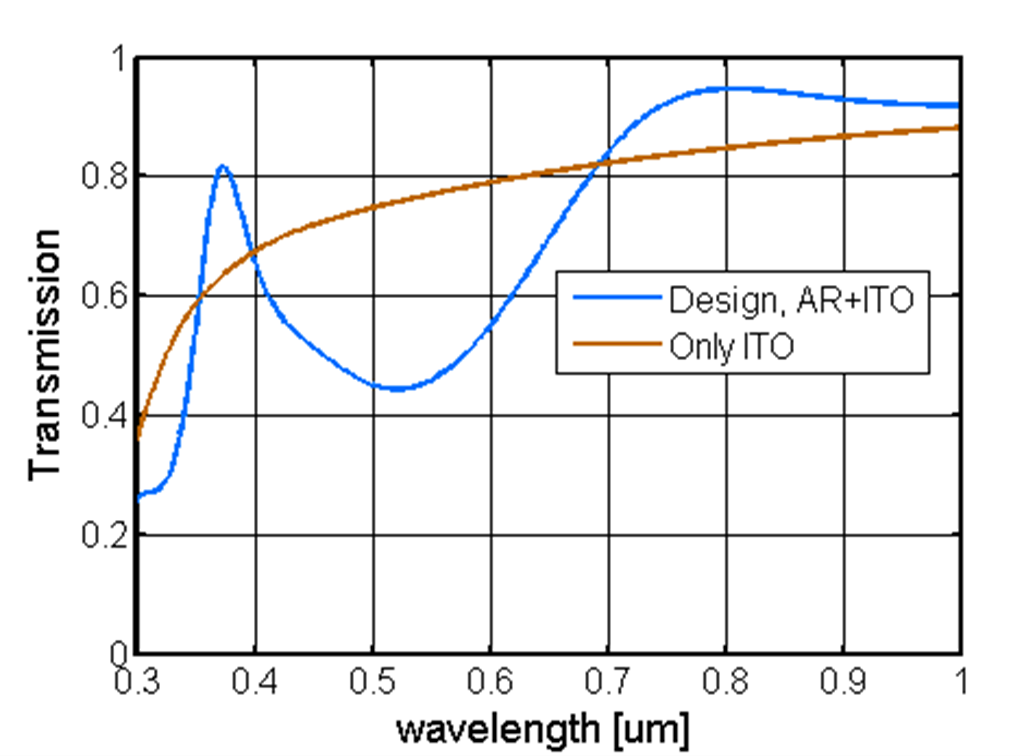}
		\label{fig: T_AR_Z}	
	\end{subfigure}
	\begin{subfigure}[h]{0.5\linewidth}
		\subcaption{}
		\includegraphics[width=\linewidth]{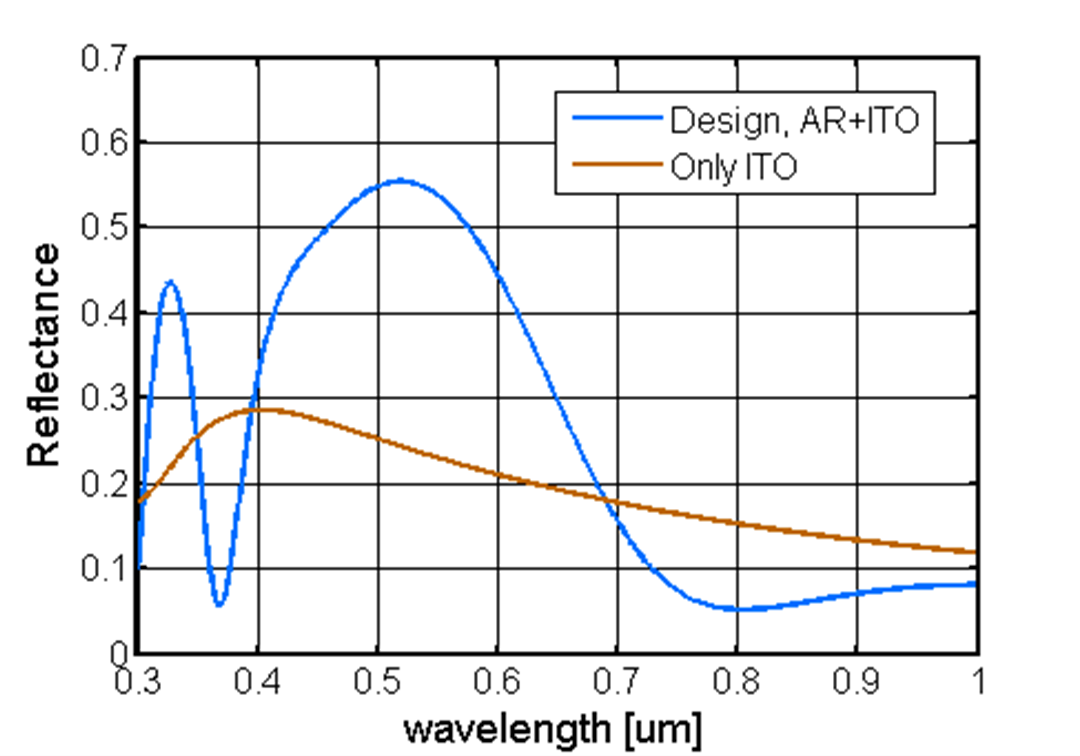}
		\label{fig:R_AR_Z}	
	\end{subfigure}
	\caption{Theoretical curves of the transmission and reflectance of the Z-optics and additionally of the used ITO layer without the underlying AR coating.}
	\label{fig:T_R_Z} 
\end{figure}

The application of the coatings presented in this paper is to ground GRIN-lenses integrated in an ion trap. The lenses are focusing laser light onto the ion in the trap and collecting the light from the fluorescing ions. Cooling and controlling Yb$^{+}$ ions in the trap requires laser beams at specific wavelengths and directions based on the electronic structure of Yb$^{+}$ and geometry of the trap setup. The light at 370 nm is used to drive the Doppler cooling transition in Yb$^{+}$ ions. Thus, the ions are also fluorescent at 370 nm and the light is collected for ion detection. Ionization of neutral Yb atoms requires 399 nm laser light and the light at 411 nm is used for electron shelving and precision spectroscopy. The laser light at 760 nm and 935 nm is used for repumping. From the wavelength and directional requirements three different coating systems were realized with high transmission at the wavelengths (see Table \ref{tab:wavelength_specs}).
\begin{table}[ht!]
		\centering
		\caption{\bf Wavelength specifications for the coatings of the GRIN-lenses (set-2 measurement samples)}
	\begin{tabular}{cc}
		\hline
		\textbf{Coating system}	& \textbf{Wavelengths $\left[ nm\right]$ } \\
		\hline
		X & 370 \\
		\hline
		Y & 370, 399, 411 \\
		\hline
		Z & 370, 760, 935 \\
		\hline
	\end{tabular}
	\label{tab:wavelength_specs}
\end{table}
\noindent For all design calculations, the previously determined ITO dispersion formula was used. As an example, we focus in detail on the developed coating system for the Z direction, since it is optimized for the broadest wavelength range. All three coating systems need a high transmission at 370 nm. This transmission is measured and the results are given later in this paper. For all three coating systems, additional high transmission at the requested wavelengths was achieved. Fig. \ref{fig: T_Z} shows the calculated transmission of coating system Z at a transmission angle of 0$\degree$ and 10$\degree$ with a 51 nm thick ITO coating, marked at the relevant wavelengths. Fig. \ref{fig: T_AR_Z} and \ref{fig:R_AR_Z} shows the difference in the transmission and reflection with and without an AR coating underneath the ITO layers for the Z coating system. As shown in Fig. \ref{fig:R_AR_Z}, there is a reduction of reflectance of the ITO layer around 370 nm and in the range from 700 to 1000 nm with the developed underlying AR design for the coating. Also in this case, all theoretical curves were calculated taking into consideration both interfaces of the test substrate.
\section{Selection of ITO thickness}
\begin{figure}[ht!]
	\centering
	\includegraphics[width=0.6\linewidth]{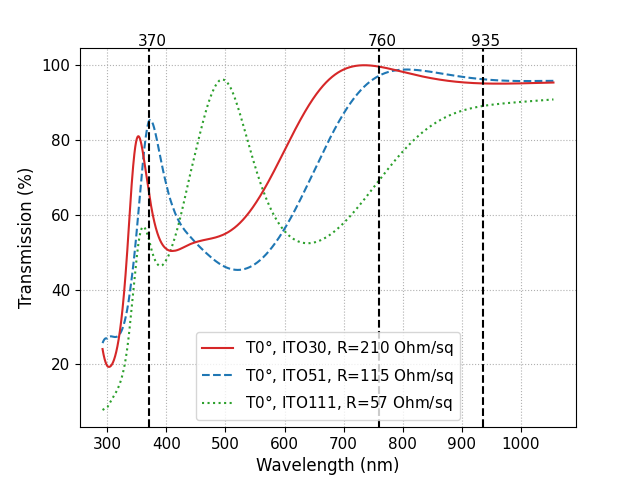}
	\caption{Comparison of the theoretical transmission curves of the Z-optics for different ITO thicknesses and their corresponding resistances.}
	\label{fig:Z_ITO_thickness} 
\end{figure}

As the conductivity of the ITO layers depends on the thickness (see Table \ref{tab:thickness_vs_resistance}), the optimal thickness of the ITO layer for all three coating systems had to be selected. For the Z-optics an ITO layer of 51 nm was chosen as the best compromise between a high enough transmission and a low enough resistance (see Fig. \ref{fig:Z_ITO_thickness}). Also for the two other optics an ITO layer of the same thickness was the best solution. 
\section{Test substrates}
\begin{figure}[ht!]
	\centering
	\includegraphics[width=0.8\linewidth]{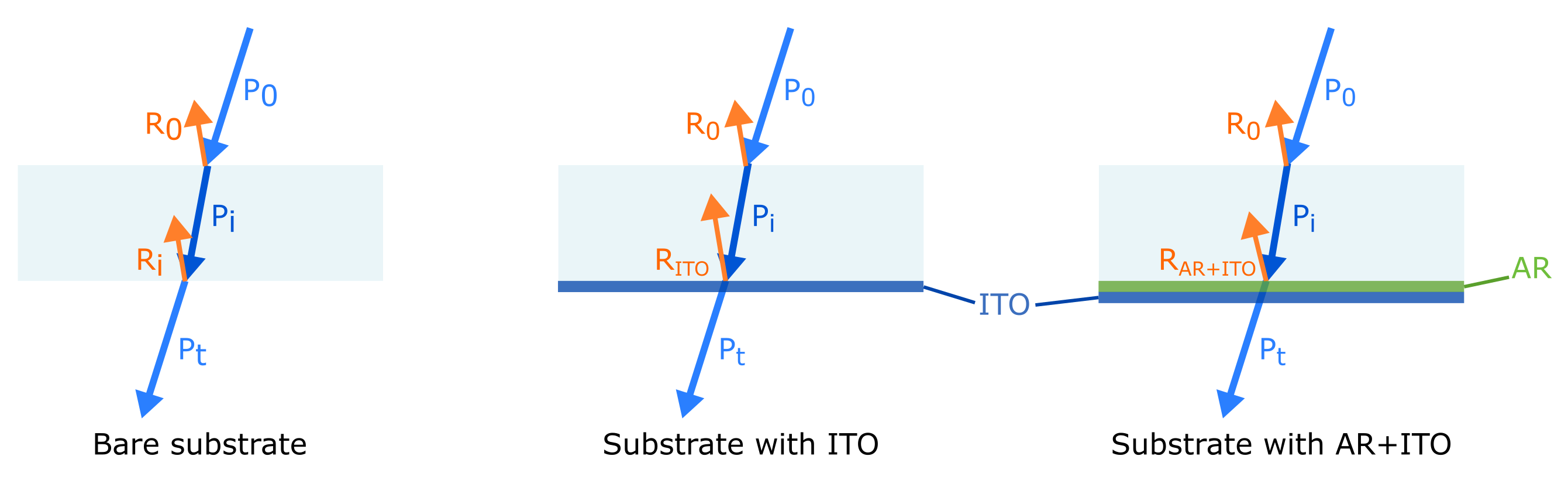}
	\caption{Various optical interfaces at the samples. $P_0$ is the incident, $P_i$ the internal and $P_t$ the transmitted power at the samples. $R_0$ is the reflectance of the incident power at the front side of the substrate, $R_i$ is the reflectance of internal power at the backside of the substrate, $R_{ITO}$ is the reflectance of internal power at the ITO layer on the backside of the substrate and $R_{AR+ITO}$ the reflectance of internal power at the AR+ITO layer on the backside of the substrate. $R_i < R_{AR+ITO} < R_{ITO}$.}
	\label{fig:Ex_optical_interfaces} 
\end{figure}

The coatings are applied on GRIN-lenses, focusing the light supported by fibers to and from the trap. Measurements of the coatings using these parts are very complicated and limited due to the small dimensions and the gradient refractive index of the GRIN-lenses. Therefore, test samples on plane, homogeneous, 1-inch diameter BK7 plates were produced in one process together with the final GRIN lenses (set-2 samples). BK7 was chosen as the substrate material, because its refractive index is very similar to that of the GRIN-lenses. The front side of the substrates (substrate-air interface) adds an additional element to the measurement set. The backside of the test substrates (substrate-air interface or substrate-coating-air interface) corresponds to the surface of the GRIN-lenses, which is the interface of interest that was characterized. The reflection and transmission at various optical interfaces of the samples are explained in Fig. \ref{fig:Ex_optical_interfaces}.
\noindent The contributions of the front side and backside of the laser transmission measurements are strongly influenced by the angle of incidence, the thickness and wedge of the substrates and the shape of the laser beam. Their exact calculation is complex, but it is a good approximation to consider that internal power (striking the backside) in all three cases equals incident power reduced by half of the measured transmission loss of the bare substrate (i.e. assuming front and back side has the same reflectance),
\begin{equation}
	P_i=P_0-\frac{(P_0-P_t)}{2},
	\label{eq:substrate_interface_power_contributions}
\end{equation}
\noindent where $P_0$ is the incident power, $P_t$ is the transmitted power and $P_i$ is the internal power. This estimation is used in Table \ref{tab:T_set1} and Table \ref{tab:T_set2}, which present the measurement results. Two sets of samples were used to characterize the coatings. In the first set of samples, ITO layers of different thicknesses (see Table \ref{tab:thickness_vs_resistance}) are coated on 1 mm thick BK7 substrates. The purpose of the different ITO thicknesses is to investigate, if the thickness affects the transmission and scattering of the coatings. In the second set of samples, 6.35 mm thick BK7 substrates were coated together with the GRIN-lenses with the AR+ITO coatings. The measurements were carried out with all three developed AR coatings X, Y and Z (see Table \ref{tab:wavelength_specs}). In all cases, a 51 nm ITO top layer with a sheet resistance of about 115 $\Omega/\Box$ (see Table  \ref{tab:thickness_vs_resistance}) was used. The surface roughness of the bare substrates was measured with an interference microscope (Micromap) (see Table \ref{tab:T_P_0}).
\section{Transmittance measurements}
\begin{figure}[ht!]
	\centering
	\includegraphics[width=0.8\linewidth]{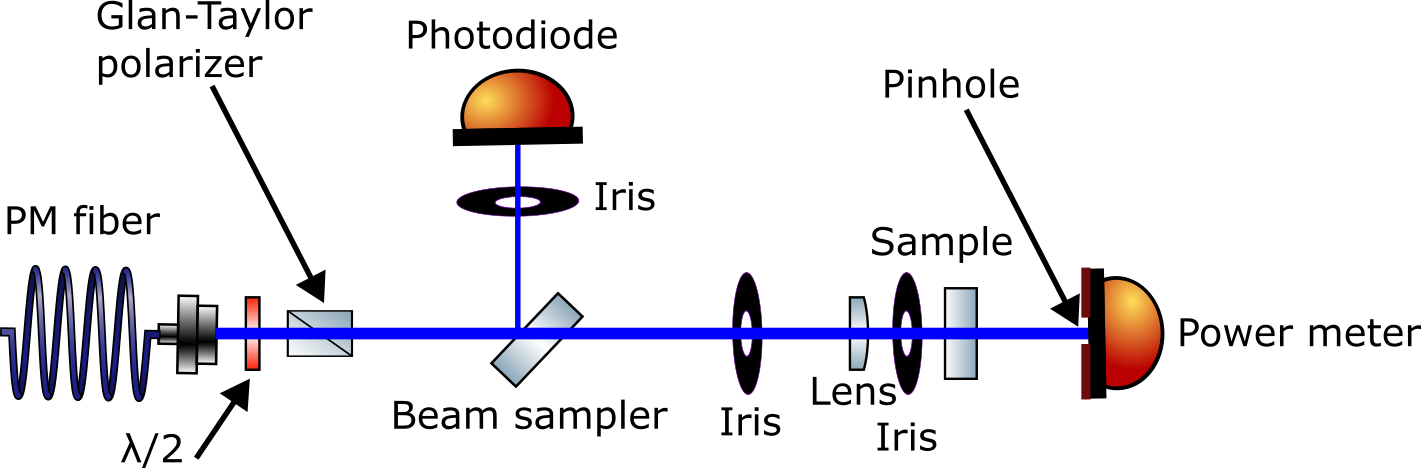}
	\caption{Sketch of the setup for the transmittance measurement.}
	\label{fig:Transmission_setup} 
\end{figure}

To measure the transmittance of the samples, a 370 nm laser was used. A sketch of the transmittance setup is shown in Fig. \ref{fig:Transmission_setup}. The laser was sent through a half-wave plate and a Glan-Taylor polarizer (polarization extinction 50 dB) to ensure good polarization stability. 2$\%$ of the power of the beam is reflected by a beam sampler onto a photodiode for monitoring the laser power, while the rest of the beam is focused to a waist of about 37 $\mu$m by a f=100 mm lens. The test sample is positioned within the Rayleigh range of the focused Gaussian beam. The laser power at the sample is about 640 $\mu$W for set-1 and about 490 $\mu$W for set-2. Apertures block the back scattered light from the setup. The transmitted light is measured by a power meter after a 1 mm pinhole.
\subsection{Substrate transmittance measurement}
The transmittance of the bare substrates was measured by comparing the laser power after the substrates to the incident laser power. The internal power was calculated using Eq. (\ref{eq:substrate_interface_power_contributions}) and the result for both substrates is given in Table \ref{tab:T_P_0}. The internal power of the set-1 substrate is lower by 2.7$\%$ than that of the set-2 substrate. From the difference in surface roughness, we expect a smaller relative internal power difference. Another reason for the difference in transmittance could be contamination of the surface, however no contamination was observed. The most likely explanation for the difference in internal power is the thickness difference between the substrates in combination with the focused incident beam. All measurements of the ITO layers on these substrates are performed under the same conditions as the measurements of the corresponding substrates. Thus the use of the internal power value should be correct. Besides, a slight change of the given transmittance values of Table \ref{tab:T_set1} and Table \ref{tab:T_set2} does not affect the main findings of our measurements.
\pagebreak
\begin{table}[ht!]
		\centering
		\caption{\bf The material properties and the coatings of the different substrates, as well as the incident power, the transmittance scaled by the laser power fluctuations and the internal power of the substrates.}
	\begin{tabular}{ccc}
		\hline
		\textbf{Item} &\textbf{Set-1} &\textbf{Set-2} \\
		\hline
		Substrate material & BK7 & BK7 \\
		\hline
		Substrate thickness [mm] & 1.00 $\pm$ 0.1 & 6.35 $\pm$ 0.1\\
		\hline
		Scratch/Dig & 20/10 & 10/5 \\
		\hline
		Measured substrate \newline roughness [nm] & 0.466 $\pm$ 0.007 & 0.307 $\pm$ 0.018 \\
		\hline
		Coating & ITO layers & AR+ITO layers\\
		\hline
		Incident power [$\mu$W] $P_0$,\newline bare substrate & 646 $\pm$ 0.006 &  492 $\pm$ 0.006 \\
		\hline
		Relative transmitted power \newline $P_t=P/P_0$, bare substrate & 0.855 $\pm$ 0.006 & 0.906 $\pm$ 0.006 \\
		\hline
		Calculated internal power\newline $P_i=P_0-\frac{(P_0-P_t)}{2}$ & 0.927 $\pm$ 0.003 & 0.953 $\pm$ 0.003 \\
		\hline	
	\end{tabular}
	\label{tab:T_P_0}
\end{table}
\subsection{Transmittance of the set-1 samples (ITO coatings)}
The transmittance of all coatings at the backside of the substrates is scaled with the internal power of the substrates and the laser power fluctuations. The transmittance of the samples with different ITO thicknesses is listed in Table \ref{tab:T_set1}. As discussed above, the transmittance of ITO is not only a function of ITO thickness but also of the optical interference between the front and back surface of the substrate. Therefore, the transmittance of sample set-1 is lower compared to sample set-2 (see Table \ref{tab:T_set2}) due to the lack of AR coating. Furthermore, the samples have an uncoated backside, which gives a slightly lower transmittance compared to the calculated transmittance in Fig. \ref{fig: T_R_diff_thickness}. 
\begin{table}[ht]
		\centering
		\caption{\bf Transmittance $\frac{P_{t}}{P_{i}}$ at 370 nm of set-1 samples for different ITO coating thicknesses, scaled with the internal power and the laser power fluctuations. The values are slightly lower compared to the simulations in Fig. \ref{fig: T_R_diff_thickness} because the backside of the substrate is uncoated.}
	\begin{tabular}{cc}
		\hline
		\textbf{sample}	& \textbf{Transmittance} \\
		\hline
		ITO 30 nm & 0.635 $\pm$ 0.004\\
		\hline
		ITO 51 nm & 0.661 $\pm$ 0.004\\
		\hline
		ITO 111 nm & 0.615 $\pm$ 0.003\\
		\hline
	\end{tabular}
	\label{tab:T_set1}
\end{table}
\subsection{Transmittance of the set-2 samples (AR+ITO coatings)}
The measured transmittance of different AR+ITO coatings are listed in Table \ref{tab:T_set2}. 
\begin{table}[ht]
		\centering
		\caption{\bf Transmittance $\frac{P_{t}}{P_{i}}$ at 370 nm of set-2 samples for coating with AR+ ITO, scaled with the internal power and the laser power fluctuations. The values are slightly lower compared to the simulations in Fig. \ref{fig: T_AR_Z} because the backside of the substrate is uncoated.}
	\begin{tabular}{cc}
		\hline
		\textbf{sample}	& \textbf{Transmittance} \\
		\hline
		X & 0.809 $\pm$ 0.005\\
		\hline
		Y & 0.788 $\pm$ 0.005\\
		\hline
		Z & 0.790 $\pm$ 0.005\\
		\hline
	\end{tabular}
	\label{tab:T_set2}
\end{table}
\noindent The samples contain an AR coating underneath a 51 nm thick ITO coating, and the measurements are similar to the simulations with around 80 $\%$ transmittance at 370 nm. The samples have an uncoated backside, which gives a slightly lower transmittance compared to the calculation in Fig. \ref{fig: T_AR_Z}. As expected, the transmittance of the samples with an AR layer underneath the ITO layer is higher than the transmittance of the 51 nm ITO layer of set-1. The X sample had the highest transmittance, because the AR coating is optimized at 370 nm. 
\section{Scattering measurements}
\begin{figure}[ht!]
	\centering
	\includegraphics[width=0.8\linewidth]{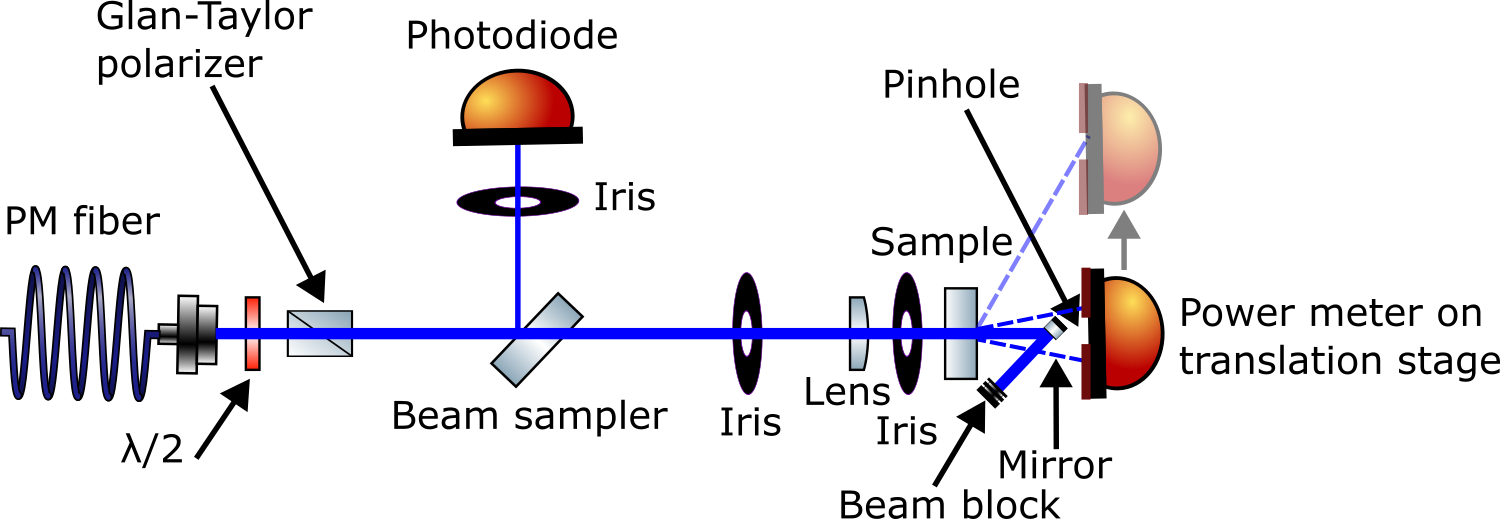}
	\caption{Sketch of the setup for the scattering measurement.}
	\label{fig:Scattering_setup} 
\end{figure}

To measure the scattered light, a small mirror (1.2 mm wide) is added to deflect the central beam onto a beam block (see Fig. \ref{fig:Scattering_setup}) shadowing a cone of about $\pm$ 1.6$\degree$. The scattered light from the samples were measured with a power meter in an angular range from 2$\degree$ to 13$\degree$. The pinhole with 1 mm diameter in front of the power meter ensures a better angular resolution. 
\subsection{Results of scattering measurements of set-1 samples with different ITO thicknesses}
\begin{figure}[ht!]
	\centering
	\includegraphics[width=0.6\linewidth]{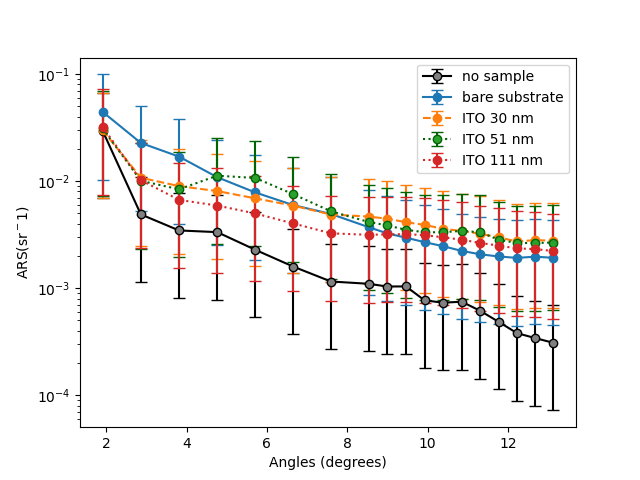}
	\caption{The angle resolved scattering as defined by Eq. (\ref{eq:ARS}) for the background without any sample (black), the bare substrate (blue) and the samples with different ITO thicknesses without AR coating described in Table \ref{tab:thickness_vs_resistance}.}
	\label{fig:ARS_diff_thickness} 
\end{figure}

First, the background without a sample, then the bare substrate and the coated samples were measured. Fig. \ref{fig:ARS_diff_thickness} shows the angle resolved scattering (ARS) defined by \cite{von_finck_instrument_2011},
\begin{equation}
	ARS(\theta)=\frac{P_s(\theta)}{\Omega P_i},
	\label{eq:ARS}
\end{equation}
\noindent where $P_s (\theta)$ is the scattered power into the solid angle $\Omega$ corresponding to the pinhole of the detector and $P_i$ is the internal power.
\noindent The three ITO layers show a similar behavior. All three layers show a scattering at the level of the bare substrate. The total scattering (TS) in the measured angular range is given by integrating the ARS curves over the measured angular range using
\begin{equation}
	TS=2 \pi \int_{\theta_{min}}^{\theta_{max}}ARS(\theta)\sin{\theta}d\theta
	\label{eq:TS}
\end{equation}
and the results are presented in Table \ref{tab:Scattering_set1}. 
\begin{table}[ht]
	\centering
	\caption{\bf The total scattered (TS) 370 nm laser light at 640 $\mu$W in the measured range for each ITO thickness. Here the substrate thickness is 1 mm and roughness is 0.466 $\pm$ 0.007 nm.}
	\begin{tabular}{ccc}
		\hline
		\textbf{Measurement}	& \textbf{TS ($\%$)} &\textbf{Background subtracted TS ($\%$)}  \\
		\hline
		No sample & 0.026$\pm$0.002 &  \\
		\hline
		Bare substrate & 0.088$\pm$0.006& 0.062$\pm$0.006  \\
		\hline
		ITO30 & 0.081$\pm$0.005 & 	0.055$\pm$0.005 \\
		\hline
		ITO51 & 0.087$\pm$0.005	& 0.061$\pm$0.006  \\
		\hline
		ITO111 & 0.063$\pm$0.003	 & 0.038$\pm$0.004 \\
		\hline
	\end{tabular}
	\label{tab:Scattering_set1}
\end{table}
\subsection{Results of scattering measurements of ITO coated samples with AR coating}
\begin{figure}[ht!]
	\centering
	\includegraphics[width=0.6\linewidth]{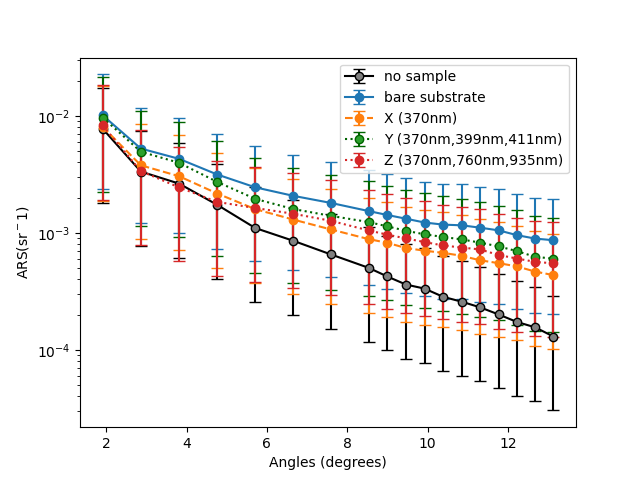}
	\caption{The figure shows the angle resolved scattering as defined by Eq. (\ref{eq:ARS}) for the background without any sample (black), the bare substrate (blue) and the samples with ITO+AR coating described in Table \ref{tab:wavelength_specs}. Here the ITO thickness is 51 nm. }
	\label{fig:ARS_IDEAL} 
\end{figure}
\noindent Fig. \ref{fig:ARS_IDEAL} shows the ARS of the background without any sample, the bare substrate without any coatings and the coated substrates described in Table \ref{tab:wavelength_specs}. Since all three coating systems show a scattering at the level of the bare substrate, we conclude that the bare substrate is the dominating source of scattering. The total scattering of the samples is presented in Table \ref{tab:Scattering_set2}. \pagebreak
\begin{table}[ht]
	\centering
	\caption{\bf The total scattered (TS) 370 nm laser light at 490 $\mu$W in the measured range for the measurements with AR+ITO coating. Here the substrate thickness is 6.35 mm and roughness is 0.307 $\pm$ 0.018 nm.}
	\begin{tabular}{ccc}
		\hline
		\textbf{Measurement}	& \textbf{TS ($\%$)} &\textbf{Background subtracted TS ($\%$)}  \\
		\hline
		No sample & 0.0123$\pm$0.0009 &  \\
		\hline
		Bare substrate & 0.029$\pm$0.002& 0.017$\pm$0.002 \\
		\hline
		X & 0.019$\pm$0.001 & 0.006$\pm$0.001 \\
		\hline
		Y & 0.024$\pm$0.001	& 0.012$\pm$0.002  \\
		\hline
		Z & 0.019$\pm$0.001	 & 0.007$\pm$0.001\\
		\hline
	\end{tabular}
	\label{tab:Scattering_set2}
\end{table}
\section{Discussion of the scattering measurement results}
\noindent The total scattering is larger for the set-1 (only ITO) compared to set-2 (ITO+AR coating). For the samples with AR coating, an ITO thickness of 51 nm was used, which is the same thickness as the sample with 51 nm ITO in set-1. Therefore scattering of the ITO is expected to be the same in both cases. The difference in the total scattering indicates that there are other factors influencing the measurement. One such factor is the influence of the AR coating on the measurement. The ARS, (defined in Eq. (\ref{eq:ARS})) depends on the internal power (defined in Eq. (\ref{eq:substrate_interface_power_contributions})) and by combining the equations we get,
\begin{equation}
		ARS(\theta)=\frac{P_s(\theta)}{\Omega P_i}=\frac{2 P_s(\theta)}{\Omega (P_0+P_t)}\propto\frac{2 P_s(\theta)}{\Omega (2P_0-R)}, 
		\label{eq:ARS_influence_on_scattering}
\end{equation}
\noindent where $R$ is the reflectance of the incident power at the surface-coating interface. Therefore, a larger reflectance of the incident power leads to a larger ARS value for the samples without AR coating. Another factor that influences the scattering is the surface roughness of the samples \cite{Hultaker_2003}. Since the samples have different surface roughness, we also expect different scattering levels. From the total scattering we can estimate the surface roughness of the substrates. Since we are measuring the forward scattering of the samples, we need to take into account the scattering from both the front and the back surface of the substrate. Assuming both surfaces have the same roughness, we get   
\begin{equation}
	TS=TS_{front}+TS_{back}=TS_{front}(1+\tau_{0}),
	\label{eq:Total_scattering_based_on_front_and_back}
\end{equation}
where $\tau_{0}$ is the transmittance of the sample. From the front scattering we obtain the following surface roughness based on reference \cite{Hultaker_2003},
\begin{equation}
	\sigma=\frac{\lambda}{2\pi(n-1)}\sqrt{\frac{TS_{front}}{\tau_{0}}}=\frac{\lambda}{2\pi(n-1)}\sqrt{\frac{TS}{\tau_{0}(1+\tau_{0})}},
	\label{eq:surface_roughness}
\end{equation}
where $\lambda$ is the wavelength and $n$ is the refractive index. We estimate the surface roughness of the substrates using the measured total scattering from Table \ref{tab:Scattering_set1} and \ref{tab:Scattering_set2} and the transmittance from Table \ref{tab:T_P_0}. The results are shown in Table \ref{tab:surface_roughness}. 
\begin{table}[htbp!]
	\centering
	\caption{\bf Surface roughness for different samples estimated based on the scattering measurements and measured using an interference microscope. The substrate thickness was 1 mm for set-1 and 6.35 mm for set-2.}
	\begin{tabular}{ccc}
		\hline
		\textbf{Sample}	& \textbf{Surface roughness (nm) from TS}& \textbf{Surface roughness (nm) measured}\\
		\hline
		bare substrate set-1 & 2.0 $\pm$ 0.1 & 0.466 $\pm$ 0.007\\
		\hline
		bare substrate set-2 & 1.05 $\pm$ 0.06 &0.307 $\pm$ 0.018\\
		\hline
		ITO 51 nm set-1 & 2.6 $\pm$ 0.1 & 0.435 $\pm$ 0.006\\
		\hline
		ITO Z set-2  & 0.77 $\pm$ 0.06 & 0.286 $\pm$ 0.011\\
		\hline
	\end{tabular}
	\label{tab:surface_roughness}
\end{table}
The estimated surface roughness based on the total scattering is larger compared to the surface roughness  measured with an interference microscope (Micromap). An explanation for this is that the measured surface roughness is the average of the roughness of the measured surface area and therefore singular peaks and valleys on the surface only have a minor contribution to the overall average. In the scattering measurement, we are measuring the scattering of the surface area with the laser spot size and therefore singular peaks and valleys can impact the measurement results. Another factor that can influence the estimate of the surface roughness is the bulk scattering \cite{Hultaker_2003}. Since the substrates of set-1 are thinner (1 mm) than the substrates of set-2 (6.35 mm) and both are BK7, we expect the bulk scattering to be smaller in set-1 compared to set-2. However, since the estimated surface roughness is larger for the thin substrate, we don't see any influence of the bulk scattering. This could be due to the surface scattering dominating the measurement or an increased bulk scattering in set-1 for example due to the material composition.  
\section{Conclusion}
We present ITO coatings in combination with AR coatings designed for optically integrated ion traps with Yb$^{+}$ ions. For controlling the Yb$^{+}$ several wavelengths between 370 nm and 935 nm are required. We measured the transmittance and optical scattering at the shortest wavelength of interest 370 nm. The measurements show that the AR coating below the ITO layer result in an increase of the transmittance through the ITO layer to about 80$\%$ at 370 nm with a resistance of 115 $\pm$ 5 $\Omega/\Box$, whereas the scattering remains very low, with the highest level of scattering of 0.012$\pm$0.002$\%$. The ARS for the bare substrates are similar to the ARS of the ITO coated samples, therefore we conclude that the main source of scattering comes from the bare substrate and not the ITO coatings. We found that the AR coating is very effective for increasing the transmittance from 66$\%$ to 80$\%$. \\
Regarding the suitability of the coatings for the ion trap with integrated GRIN-lenses, we find that the transmittance of 80$\%$ at 370 nm is sufficient for controlling the ions. Furthermore, the coatings were produced at a temperature below 150 $\degree$C, thus the coating process is suitable for the GRIN-lenses. At temperatures $> 150 \degree$C the dopand in the GRIN lenses can diffuse and the lenses get degraded. Although the scattering is low and the main source of scattering is the substrates, it is still necessary to investigate how the scattering would affect an ion trap experiment. There are two main ways that the scattering can affect the performance of an ion trap experiment. The first one is by crosstalk for the addressing of ions in different trapping zones and the second is by increasing the detection noise. Assuming a distance of 10 mm between the lens and the ion and an angle of 13°. This corresponds to a distance between ions in different trapping zones of about 2 mm. For a hot ion, with a spacial distribution of 1 µm in diameter we get a scattering at the ion of $5\cdot 10^{-8}\%$, which corresponds to a suppression of about 73 dB. To estimate the detection noise caused by the scattering, we check if the scattering from the ITO+AR coated GRIN-lenses can enter the image created by the imaging system. In our case, the scattered light for the X, Y and Z direction is never part of the images generated by the detection lenses because of the trap design. \\
We conclude that these coatings are suitable for ion trap experiments. The achieved good conductivity of the ITO coating protects the surfaces from charging which could otherwise lead to a shift in the ions position, increased micromotion and anomalous heating. The high transmittance in the UV and the low scattering allows for efficient laser cooling and reliable state readout. Although the scattering is low and mainly from the substrates, it can have an impact on the crosstalk between ions in different trapping zones. Therefore, it is important to take that into account when designing an ion trap experiment with integrated optics. 
\section*{Funding and acknowledgements}
This work was supported by the BMBF (Federal Ministry of Education and Research, Germany) Grant number 13N14959, ’Integrierte Diamant Ionenfalle (Integrated diamond ion traps) - IDEAL’. Erik Jansson acknowledges support from the Max-Planck-RIKEN-PTB-Center for Time, Constants and Fundamental Symmetries and the Deutsche Forschungsgemeinschaft (DFG, German Research Foundation) under Germany’s Excellence Strategy – EXC-2123 QuantumFrontiers – 390837967. The authors thank Andr\'{e} Felgner for the surface roughness measurements and related discussions and Daniel Kopf for the support and discussions of the measurement results. 
\section*{Disclosures}
The authors declare no conflicts of interest.
\section*{Data Availability Statement}
The original data is included in the article, further inquiries can be directed
to the corresponding author.
\section*{Supplemental material} See Supplement 1 for supporting content.

\end{document}


\maketitle

\section{Transmittance measurement reproducability}
The transmittance of sample set-2 was measured on three different days at different rotations of the sample. The results are listed in Table \ref{tab:T_set2_reproducability}.\\
\begin{table}[ht]
	\centering
	\caption{\bf Transmittance of 370 nm of set-2 samples (coating with AR+ ITO) measured on three different days at different rotations of the sample.}
	\begin{tabular}{cccc}
	\hline
	\textbf{Sample}	& \textbf{Measurement 1} &  \textbf{Measurement 2} & \textbf{Measurement 3} \\
	\hline
	Substrate & 0.955 $\pm$ 0.004 & 0.953$\pm$ 0.003 & 0.956 $\pm$ 0.004 \\ 
	\hline
	X & 0.805 $\pm$ 0.006 & 0.809 $\pm$ 0.005 & 0.808 $\pm$ 0.006\\
	\hline
	Y & 0.792 $\pm$ 0.005 & 0.788 $\pm$ 0.005 & 0.781 $\pm$ 0.005\\
	\hline
	Z & 0.794 $\pm$ 0.005 & 0.790 $\pm$ 0.005 & 0.790 $\pm$ 0.005\\
	\hline
	\end{tabular}
	\label{tab:T_set2_reproducability}
\end{table} \\
\noindent The transmittance is averaged over 30 s with one measurement per second. The uncertainties comprise the uncertainty in the calibration of the photodiode, the uncertainty of the photodiode and the uncertainty of the power meter. The uncertainty of the power meter of 2.4$\%$ is the largest contribution to the overall uncertainty of the measurement.
\section{Scattering measurement reproducability}
\begin{figure}[ht!]
\centering
\includegraphics[width=\linewidth]{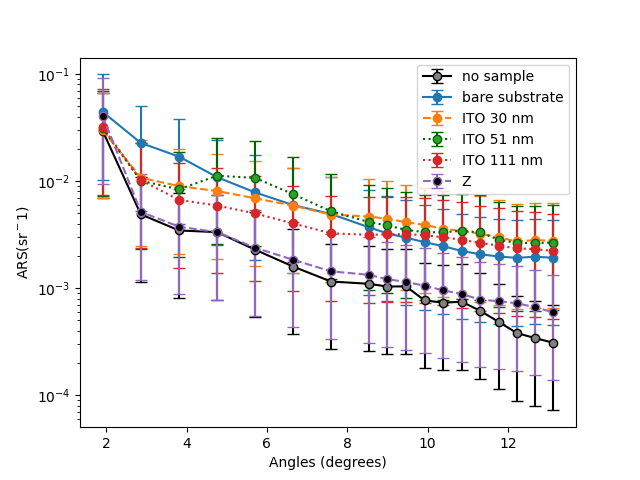}
\caption{The angle resolved scattering for the background without any sample (black), the bare substrate (blue), the samples with different ITO thicknesses and the Z sample. The bare substrate and the samples with different ITO thicknesses has a substrate thickness of 1 mm and a surface roughness of 0.466 $\pm$ 0.007 nm, while the Z sample has a substrate thickness of 6.35 mm and a surface roughness of 0.307 $\pm$ 0.018 nm and contains an AR coating as well as an ITO coating of 51 nm.}
\label{fig:ARS_diff_thick_Z}
\end{figure}

\begin{figure}[ht!]
\centering
\includegraphics[width=\linewidth]{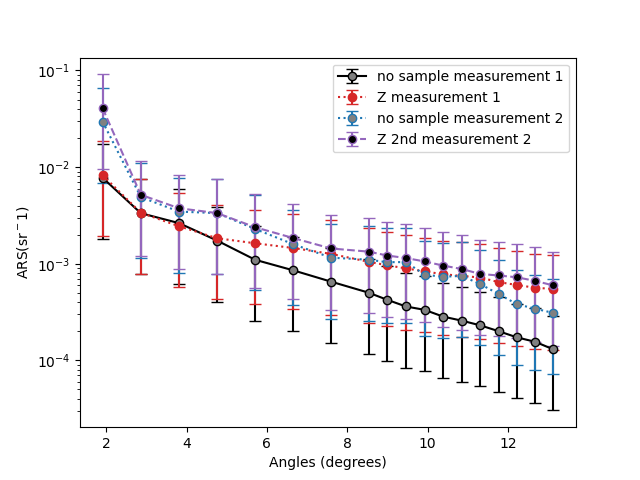}
\caption{The angle resolved scattering for two separate measurements for the background without any sample and for the Z sample. The measurement are consistent with each other, confirming the reproducability of the measurement.}
\label{fig:ARS_Z_comparisons}
\end{figure}

The Z sample was measured a second time together with the samples of different thicknesses to compare the two measurements. Fig. \ref{fig:ARS_diff_thick_Z} shows the result. The scattering from the Z sample is lower than the scattering from the samples of different thicknesses, confirming the results presented in the main article. Fig. \ref{fig:ARS_Z_comparisons} shows the two Z measurements. We see that both Z curves are consistent with one another, which confirms the reproducability of the measurements.
